\documentclass{cimento}

\usepackage{graphicx} 
\usepackage[]{units}
\usepackage{acronym}
\usepackage{upgreek}
\usepackage{caption}
\usepackage{subcaption}
\usepackage{ragged2e}
\usepackage{multirow}
\usepackage{float} 
\usepackage{booktabs}

\title{Measurement of fundamental physical quantities in the framework of the Lab2Go project}
\author{F.~Casaburo\from{ins:Sapienza}\from{ins:Roma1}\thanks{Corresponding author.},
N.~Marcelli\from{ins:Tor_Vergata}\from{ins:Roma2},
M.~Sorbara\from{ins:Tor_Vergata}\from{ins:Roma2}, M.~Agostinelli\from{ins:Trafelli}, P.~Astone\from{ins:Roma1}, F.~Baldassarre\from{ins:Trafelli}, F.~Brunori\from{ins:Trafelli}, S.~Crisci\from{ins:Trafelli}, G.~De Bonis\from{ins:Roma1}, X.~De Lucia\from{ins:Trafelli}, D.~De Pedis\from{ins:Roma1}, G.~De Valeri\from{ins:Trafelli}, G.~Di Sciascio\from{ins:Roma2}, R.~Faccini\from{ins:Sapienza}\from{ins:Roma1}, J.~Falato\from{ins:Trafelli}, V.~Fraietta\from{ins:Trafelli}, C.~Gatto\from{ins:Trafelli}, S.~Guadagnini\from{ins:Trafelli}, V.~Oliviero\from{ins:Trafelli}, G.~Organtini\from{ins:Sapienza}\from{ins:Roma1}, V.~Passamonti\from{ins:Trafelli}, F.~Piacentini\from{ins:Sapienza}\from{ins:Roma1}, N.~Ruggiero\from{ins:Trafelli}, M.~Salerno\from{ins:Trafelli}, S.~Sarti\from{ins:Sapienza}\from{ins:Roma1}, 
        \atque
L.~Tedesco\from{ins:Trafelli}}

\instlist{\inst{ins:Sapienza} Dipartimento di Fisica, Universit\`a di Roma - Roma, Italy
\inst{ins:Roma1} INFN, Sezione di Roma - Roma, Italy
  \inst{ins:Tor_Vergata} Dipartimento di Fisica, Universit\`a di Tor Vergata - Roma, Italy
  \inst{ins:Roma2} INFN, Sezione di Roma 2 - Roma, Italy
  \inst{ins:Trafelli} ITIS e Liceo Scientifico "L. Trafelli" - Nettuno, Italy
  }

\begin{document}

\maketitle

\begin{abstract}
To establish a closer contact between school and experimental sciences, Sapienza Università di Roma and the \ac{INFN} launched the Lab2Go project. Lab2Go has the goal of spreading laboratory practice among students and teachers in high schools. In this article, it is presented a measurement, carried out in the framework of the Lab2Go project, of the ratio $\frac{hc}{e}$ where $h,\,c$ and $e$ are respectively the Planck's constant, the speed of light in the vacuum,  and the electric charge.
\end{abstract}

\section{Introduction}
\label{Sec:Introduction}
Modern physics is based on a few key discoveries made across the 19th and 20th centuries: the invariance of speed of light $c$ in different reference frames (Michelson-Morley experiment \cite{MM-experiment}), the quantisation of the electric charge (Millikan experiment \cite{Millikan-experiment}) and the quantisation of the electromagnetic waves energy through the Planck relation \cite{Passon_2017}:

\begin{equation}
E=h\nu
\label{eq:Planck_law}
\end{equation}

Nowdays, three fundamental constants $c$ (speed of light), $e$ (electron charge) and $h$ (Planck constant) related to these discoveries define (since May 2019 \cite{doi.org/10.1002/andp.201800308}) the measuring units of length, charge and mass in the \ac{SI} \cite{BureauInternationaledesPoidsetMeasures}. These three constants are involved in the relations linking measurable quantities in the \acp{LED}. \acp{LED}s are semiconductor devices emitting light at a given wavelength $\lambda$ when powered by a potential difference above a  minimum threshold value $V_g$ (gap voltage). The minimum energy needed to turn on a \ac{LED} is given by:

\begin{equation}
E=eV_g
\label{eq:Egap}
\end{equation}

Current $i$ flowing into the \ac{LED} as a function of the potential difference $V_{LED}$ (characteristic curve) has an exponential trend, but the region of the curve over the gap voltage can be approximate as a linear function. 
To estimate gap voltage $V_g$, we can interpolate data of the approximately linear region of the characteristic curve by the linear function $i\left(V_{LED}\right)=mV_{LED}+q$ and calculate the intersection between the $V_{LED}-$axis and the fit function. Then $V_g$ is given by: 
\begin{equation}
V_{g}=-\frac{q}{m}
\label{eq:VgFit}
\end{equation}
Lastly, by combining the eqs. \ref{eq:Planck_law} and \ref{eq:Egap} and solving for $V_g$, we get the relation between gap voltage $V_g$,  Planck's constant $h$, speed of light in vacuum $c$, electric charge $e$, and wavelength of the \ac{LED} $\lambda$ \cite{doi:10.1119/1.2350479}:
\begin{equation}
V_{g}\left(\frac{1}{\lambda}\right)=\frac{hc}{e}\frac{1}{\lambda}
\label{eq:Vgap}
\end{equation}
In this article we present the measurement of physical quantity $k=\frac{hc}{e}$ made in the framework of the Lab2Go  project \cite{andreotti2021il}. 

\section{Experimental setup and procedure}
\begin{figure}[htbp]
\begin{subfigure}{.5\textwidth}
  \centering
  \includegraphics[height=.66\linewidth]{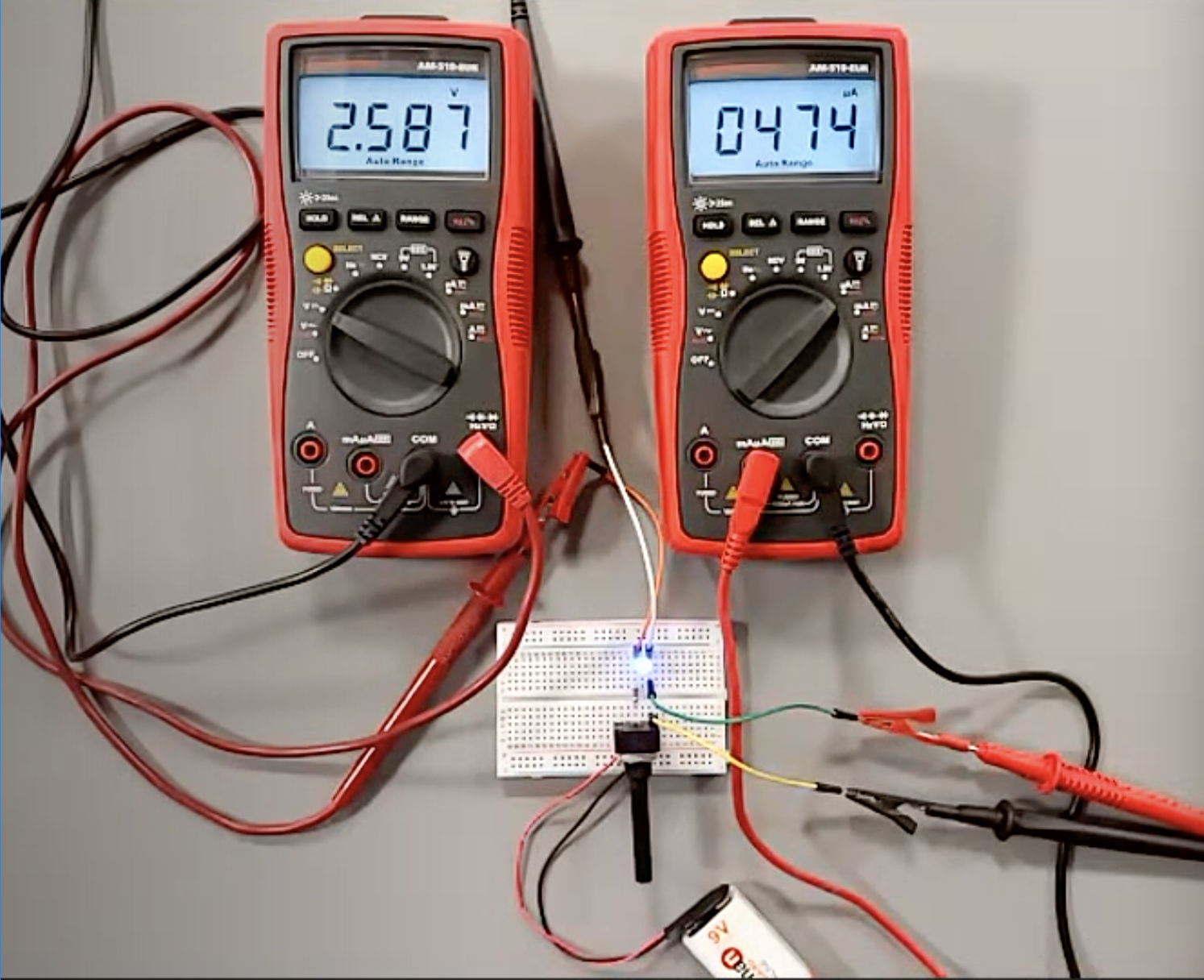}  
  \caption{}
  \label{fig:setup_analog}
\end{subfigure}
\begin{subfigure}{.5\textwidth}
  \centering
  \includegraphics[height=.67\linewidth]{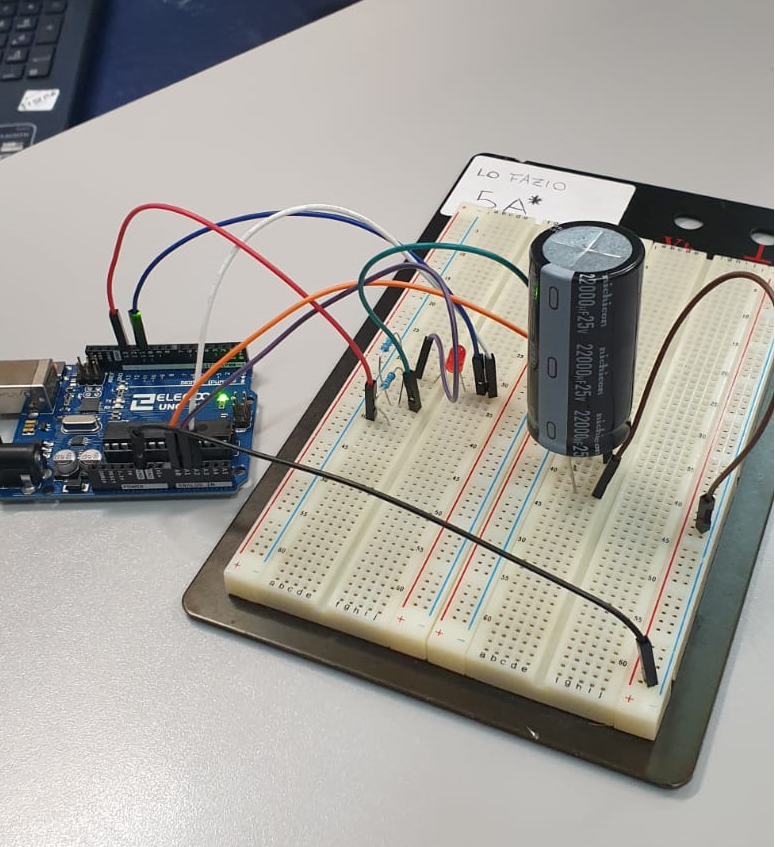}  
  \caption{}
  \label{fig:setup_arduino}
\end{subfigure}
\caption{Analog (Fig. \ref{fig:setup_analog}) and Arduino-based (Fig. \ref{fig:setup_arduino}) experimental setups.}
\label{fig:ExperimentalSetup}
\end{figure}

We present here two different approaches to perform the measurement of $k$ with instruments and methods available in secondary schools. In the first one (analog approach), \ac{LED}s are powered by a battery, the potential difference is varied by a potentiometer, and data of both the potential difference and the current are measured by two multimeters (Fig. \ref{fig:setup_analog}). In the second version (Arduino-based approach) a \unit[22]{mF} capacitor, and an Arduino board \cite{ArduinoSite} are used. Arduino is used both to charge/discharge the capacitor, and for \ac{DAQ} purposes (Fig. \ref{fig:setup_arduino}). The code (sketch) for Arduino starts charging the capacitor. When the capacitor is fully charged, by an automatic switch, it is discharged powering the \ac{LED} \cite{h_FnS}. In both versions, data of the current as a  function of the powering potential difference have been stored for several \ac{LED}s colors (red, yellow, green, and blue), corresponding to different wavelengths.  

\section{Data analysis and results}
\begin{figure}[htbp]
\begin{subfigure}{.5\textwidth}
  \centering
  \includegraphics[height=.60\linewidth]{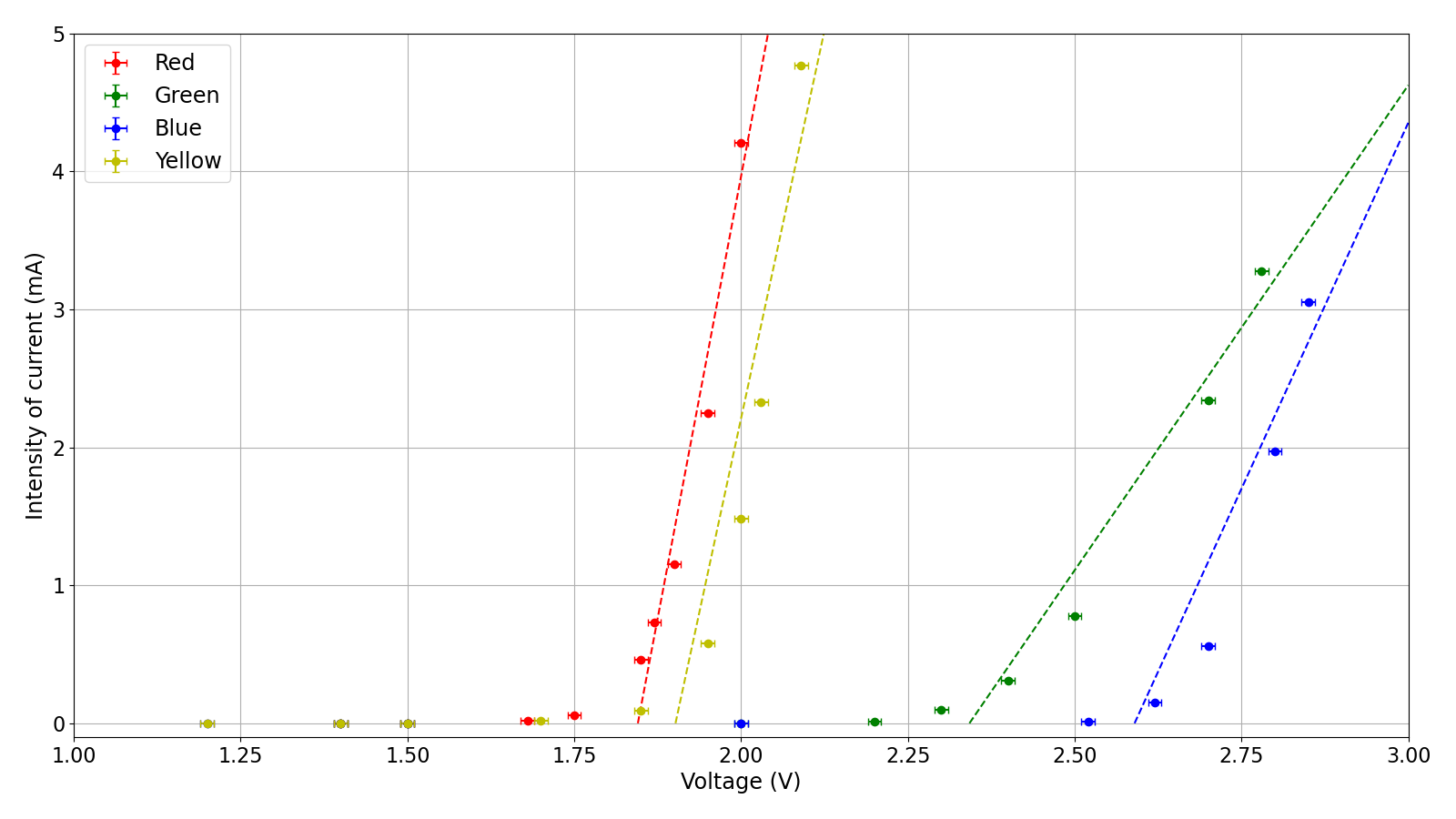}  
  \caption{}
  \label{fig:CharacteristicCurveAnalog}
\end{subfigure}
\begin{subfigure}{.5\textwidth}
  \centering
  \includegraphics[height=.67\linewidth]{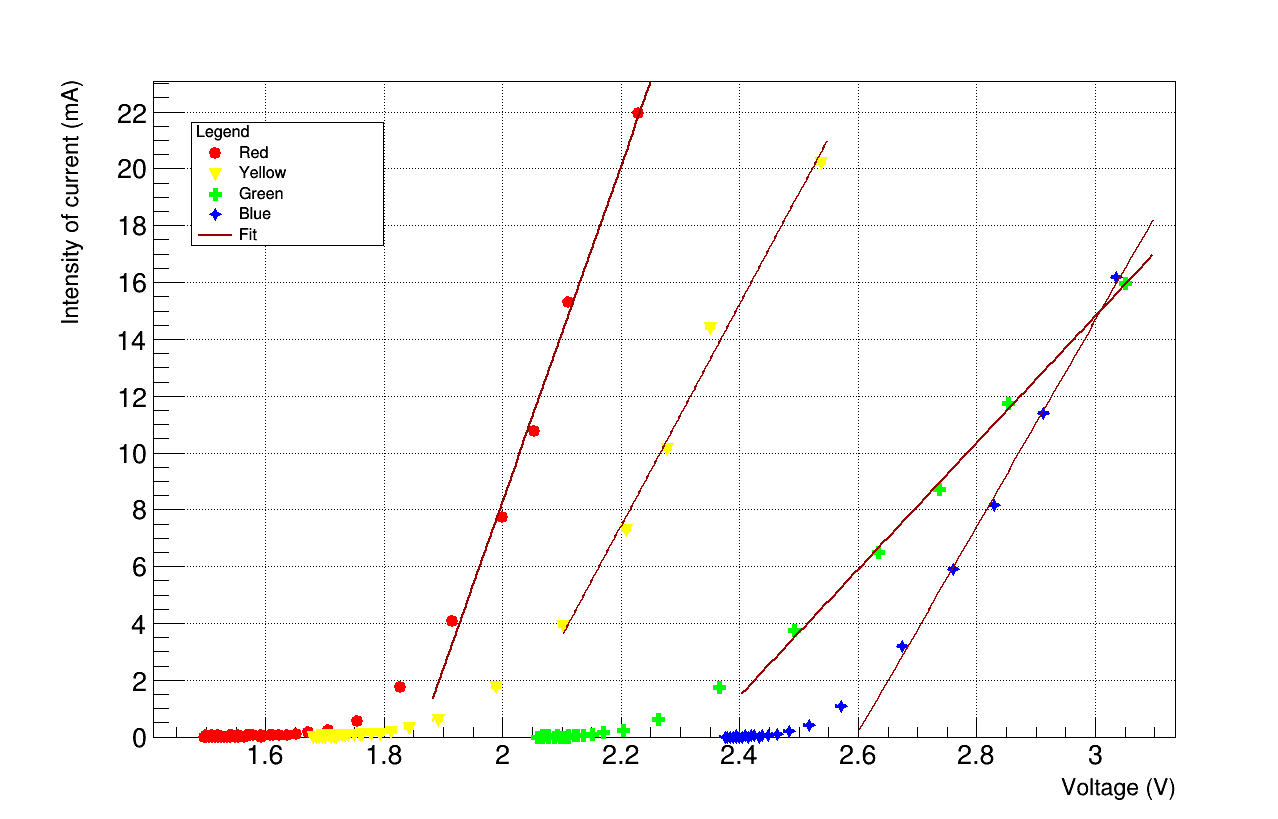}  
  \caption{}
  \label{fig:CharacteristicCurveArduino}
\end{subfigure}
\caption{Characteristic curves of \ac{LED}s obtained by the analog (Fig. \ref{fig:CharacteristicCurveAnalog}) and Arduino-based (Fig. \subref{fig:CharacteristicCurveArduino} methods. From left: red, yellow, green, and blue \ac{LED}s.}
\label{fig:CharacteristicCurves}
\end{figure}

\begin{figure}[htbp]
\centering
\includegraphics[width=3in]{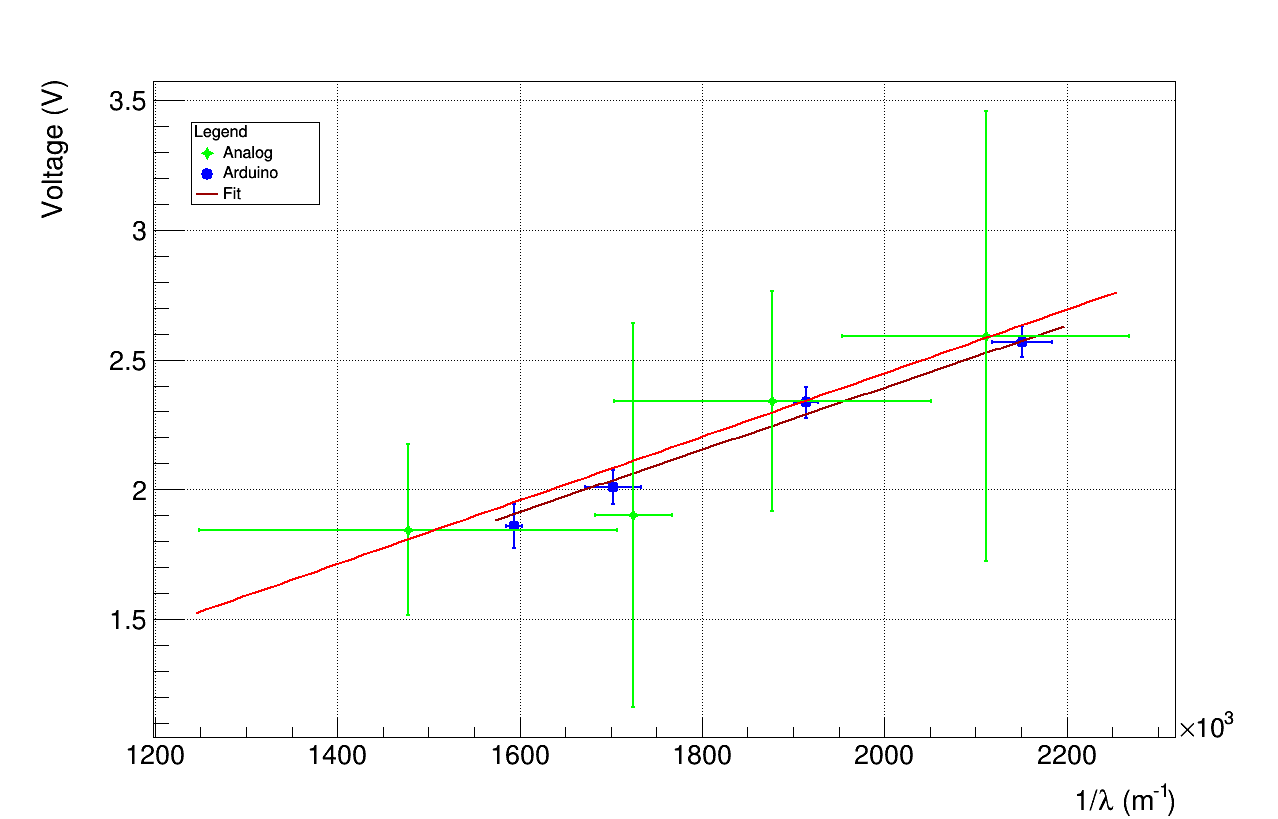}\DeclareGraphicsExtensions.
\caption{Interpolation of data of $V_g$ as a function of $\lambda^{-1}$, both for analog (green, on the top), and Arduino-based (blue, on the bottom) methods.}
\label{fig:Vg_lambda}
\end{figure}

 Data of current flowing into the \ac{LED}s as a function of the potential difference have been interpolated (Fig. \ref{fig:CharacteristicCurves}), as described in Section \ref{Sec:Introduction}, allowing estimating values of the gap voltages that are summarized in Table \ref{tab:GapTensions}. The higher uncertainties on the values of $V_g$ obtained by the analog method (Fig. \ref{fig:CharacteristicCurveAnalog}) with respect to the Arduino-based one (Fig. \ref{fig:CharacteristicCurveArduino}), just depend on data not fully contained into the fitting line. In turn, the gap voltages (Tab. \ref{tab:GapTensions}) have been interpolated (Fig. \ref{fig:Vg_lambda}) by eq. \ref{eq:Vgap} as a function of $\lambda^{-1}$ using values of the wavelengths declared in the datasheets of the \acp{LED}.
\begin{table}[htbp]
\centering
  \caption{Gap voltages of \ac{LED}s $V_g$. \label{tab:GapTensions}}
  \begin{tabular}{lllll}
 \toprule
    Method & $\unit[V_{g_{Red}}]{\left(V\right)}$ & $\unit[V_{g_{Yellow}}]{\left(V\right)}$ & $\unit[V_{g_{Green}}]{\left(V\right)}$ & $\unit[V_{g_{Blue}}]{\left(V\right)}$
    \tabularnewline
\midrule
Analog & $\ensuremath{1.85\pm0.33}$ & $\ensuremath{1.90\pm0.74}$ & $\ensuremath{2.34\pm0.42}$ & $\ensuremath{2.59\pm0.87}$\tabularnewline
Arduino & $\ensuremath{1.859\pm0.083}$ & $\ensuremath{2.009\pm0.067}$ & $\ensuremath{2.335\pm0.060}$ & $\ensuremath{2.569\pm0.059}$\tabularnewline
    \bottomrule
  \end{tabular}
\end{table}

The slope resulting by interpolation represents the estimated value of parameter $k$. Results are $\unit[k_{Analog}=\left(1.22\pm0.16\right)\cdot10^{-6}]{Vm}$ and $\unit[k_{Arduino}=\left(1.197\pm0.019\right)\cdot10^{-6}]{Vm}$ respectively for the analog and the Arduino-based methods. These values agree respectively within approximately $1\sigma$ and $2\sigma$ with the theoretical value value of approximately  $\unit[1.23984\cdot10^{-6}]{Vm}$ calculated by the accepted values of $h,\,e$ and $c$ \cite{BureauInternationaledesPoidsetMeasures}.

\section{Conclusions}
During the school year 2020/21, in the framework of Lab2Go, the measurement of fundamental physical quantities using \acp{LED} has been proposed to high school students. Despite the caveat of Planck's constant being a fundamental constant since 2019 used as a reference to define other units of measurement and for the calibration of instruments, this experiment can not be made in the future using new instruments; this proposed activity has the advantage of illustrating "quantum mechanics in action" to students.

\acknowledgments
The authors acknowledge Mauro Mancini, Francesco Safai Therani, and \ac{CC3M}-\ac{INFN}.

\bibliographystyle{ieeetr}
\bibliography{sample}

\acrodef{CC3M}[CC3M]{Comitato di Coordinamento III missione}
\acrodef{DAQ}[DAQ]{Data AcQuisition}
\acrodef{LED}[LED]{Light Emitting Diode}
\acrodef{INFN}[INFN]{Istituto Nazionale di Fisica Nucleare}
\acrodef{PLS}[PLS]{Piano Lauree Scientifiche}
\acrodef{SI}[SI]{International System of units}
\acrodef{USB}[USB]{Universal Serial Bus}

\end{document}